\documentclass[aps,preprint]{revtex4}%
\usepackage{amsmath}
\usepackage{amsfonts}
\usepackage{amssymb}
\usepackage{graphicx}

\begin{document}
\title{ Solitons and Their Arrays: from Quasi One-Dimensional Conductors to Stripes.}
\author{S. Brazovskii$^{1,2}$}
\affiliation{$^{1}$LPTMS-CNRS, UMR 8626, Univ. Paris-Sud bat. 100, Orsay,
F-91405 \footnote{brazov@lptms.u-psud.fr}}
 \affiliation{$^{2}$Landau Institute for theoretical Physics, Moscow, Russia.}

\begin{abstract}
We suggest a short review of literature on various solitonic lattices and
individual solitons in quasi one-dimensional conductors. This information seems
to be quite relevant to topics of stripes and their melted phases
correspondingly. We shall quote also the latest experiments, which access
solitons as elementary excitations in organic conductors and in charge density
waves. We shall outline a theory for ordered phases, where solitons should
acquire  forms of combined topological configurations (kink-roton complexes).
The extension of this picture to cuprates allows interpretation the latest STM
observations on local rod-like structures.
\newline\textbf{keywords}: Soliton, Confinement, Spinon, Holon,
Super-lattice, Dislocation, Topological Defect.
\bigskip
\\
Proceedings of the International Conference Stripes 06, Rome, Italy.
\newline
Journal of Superconductivity and Novel Magnetism 2007, to be publ.
\end{abstract}
\maketitle

\bigskip
\textbf{I. Solitonic lattices versus stripes.}

Inhomogeneous states of electronic systems can be viewed as a phase
separation stabilized by long range forces
\cite{nagaev,larkin-khm,rome}. The most common form is a domain
structure, alternating between different equivalent ground states
\cite{tranquada,bianconi,mook}. This is what makes the stripes to be
generically related to superstructures like solitonic lattices,
commonly existing or being expected in quasi-1D conductors. The
notion of a single soliton, as a stable elementary particle having a
finite energy, appears in the 1D limit; here the soliton is a domain
wall separating segments with different, while equivalent, ground
states.

Since more than 3 decades of its development, the science of
solitons has accumulated a vast amount of theoretical and
experimental results. Only a slice of this notion has been touched,
when theories of stripes have emerged
\cite{machida,poilblanc,schultz,zaanen}, and there was not much
counterflow as well (as a cross-fields example, quote
\cite{matv:00}). Recall the chain of conferences on "Electronic
Crystals" (see \cite{ecrys:05} for the latest event) which aims to
relate all topics of electronic aggregation in solids. The goal of
this short review is to give a reference guide to theory, with
excursions to latest experiments, of solitonic lattices and
individual solitons, and to outline extensions to general strongly
correlated electronic systems.

\bigskip
\textbf{I.1. Regular superstructures.}
\medskip

Flexibility of spontaneously formed states with a broken symmetry
allows for their local modifications when electrons are added via
doping, tunneling, photoemission, optics, or field effect. Lattices of
solitons usually appear because of incommensurability, which itself
can be derived from external or internal transfer of charge or of
spin. The oldest, phenomenological version is a case of weak
interaction between the underlying lattice and the superstructure.
Its natural form is a sin-Gordon theory, firstly suggested
\cite{dzyal} for helicoidal antiferromagnets (AFM) - spin density
waves (SDW). A similar description is allowed for interesting cases
of a mutual commensurability of coexisting superlattices, like the
lock-in of two Charge Density Waves (CDW) observed in
$\mathrm{NbSe_3}$ \cite{ayari}.

Strength and stability of solitons are particularly high for the
case of one electron per unit cell, i.e. a nearly half-filling of
the bare zone. The interest to this case was boosted by studies of a
doped polymer, polyacetylene (see reviews \cite{polymers,Yu-Lu}),
which ground state shows a spontaneous bond dimerization - the
Peierls effect. Exact solutions for periodic lattices of solitons
(see the review \cite{SB:84}) were allowed for all basic physically
realistic models. They include: two principle limits of electronic
concentration (small and nearly half \cite{BGK:80} band fillings);
arbitrary filling with effects of breaking charge conjugation
symmetry \cite{BM:84}; Zeeman spin splitting \cite{BDK:82,BM:84};
combined effect of doping and spin splitting \cite{BDK:82};
interchain hybridization \cite{BGS:82,BGL:82}; effects of lattice
discreteness \cite{BDKrich:82,BD:85}; case of superconductivity
\cite{buzdin}. Applications included: conducting polymers,
spin-Peierls chains, CDWs and SDWs, FFLO phase in superconductors.
The common spectral property: emergence of intragap bands partially
filled by electrons, is the most important feature and the origin of
energetic stability of solitonic lattices.

Typically, the solitons were found to be domain walls with exactly $\pi-$
shifted (acquiring half a period) profile and anomalous quantum numbers: they
are either charged or spin polarized. Both of these properties are consequences
of the charge conjugation symmetry: its lifting due to momentum dispersion of
interactions makes the phase shift arbitrary and brings to life a coexisting
charge and spin density \cite{BM:84}. The charge and the phase shift are not
ultimately bound, as it is shown by the richest case of doubly periodic
solitonic lattice originated by a combined effect of charge doping and spin
polarization \cite{BDK:82}. Here, the secondary solitonic superlattice
appearing above a threshold magnetic field, evolves drastically as a function
of doping: from the lattice of spin carrying polarons embedded to the lattice
of charged kinks to the overlapping superstructure of amplitude solitons. The
electric charge associated to each spin varies from nearly $e$ to nearly $0$,
but the phase shift of spin-soliton superlattice stays exactly at $\pi$.

Taking into account the interchain hybridization of electrons, i.e. the
curvature of their Fermi surfaces,  is an important step toward higher
dimensional systems, particularly the $\mathrm{CuO_2}$ planes. In quasi-1D
case, a strong band curvature can provoke formation of a neutral solitonic
lattice, even with no doping or spin polarization \cite{BGS:82,BGL:82}:
split-off intragap bands provide the necessary energy gain. Recently, this
mechanism was invoked \cite{gorkov:05} to explain the SDW-metal transition in
organic conductors under pressure. Finally, recall the solitonic structures in
quasi-1D superconductors \cite{buzdin}, which appear as a 1D version of the
well-known FFLO inhomogeneous state near the pair-breaking limit. Being very
weak in 3D, this effect becomes quite pronounced in systems with nested Fermi
surfaces which is the case of the 1D limit.

\bigskip
\textbf{I.2. Inhomogeneity and melting.}
\medskip

The above classification implied that solitonic lattices are plane
structures - the regular stripes. More complicated, and attracting
much modern interest, is the case of an inhomogeneous doping,
realized e.g. in the FET geometry, when the solitons' chemical
potential $\mu$ contains a variable electric potential
$\mu\Rightarrow2e\Phi(\vec{r})+cnst$, to be determined
self-consistently. Then, the solitonic lattice will change its
on-chain $x$- period in, say $y$ direction, while the field is
penetrating inside the host crystal $y>0$ or between regions of
varying dopant concentration. This regime results in a branching
structure of dislocations of the solitonic lattice
\cite{bm-surface,kirova-ecrys:05}.

In 1D, the solitonic lattice cannot possess a long range order: it
becomes a liquid preserving the local periodicity in its correlation
function. But what happens for a weak interchain interaction,
competing with quantum or thermal fluctuations, which question is
most relevant to stripes' melting? Details are known at least for
the gas of kink-solitons in systems with a double degeneracy (nearly
half filling) \cite{TB-SB:83}. As a function of temperature, the
system experiences two consecutive phase transitions at $T_1>T_2$
(in D=3, while in D=2 the lower $T_2$ becomes a crossover). At
highest $T>T_1$, the kinks are decoupled, serving as quasi-particles.
Below the upper "confinement transition", at $T_1>T>T_2$, the kinks
and anti-kinks are bound into neutral pairs, which may be viewed as
nucleolus droplets in magnetic semiconductors \cite{nagaev}. Below
the second transition of "aggregation", at $T<T_2$, a fraction of
pairs is broken again, and the unpaired kinks assemble into domain
walls crossing the sample. In case of charged solitons, this picture
may be essentially affected by long range Coulomb forces
\cite{teber}, which e.g. lead to instability of plane solitonic
lattices even within the coherent low $T$ phase. It may not be an
accidence, in this respect, that with a high experimental precision
profiles of solitons have been obtained (by the NMR \cite{cugeo})
only for spin chains, thanks to their high 3D coherence.

\bigskip
\textbf{II. Solitons as quasi-particles.}
\medskip

Macroscopic instability of electronic systems to formation of
solitonic lattices  can manifest itself already at the
single-particle level. The very character of elementary excitations
can be modified; they acquire forms of topological configurations -
solitons exploring the possibility to travel among different
allowed ground states. These excitations can be also viewed as
nucleuses of the melted stripe phase, which is typically observed
under higher doping of oxides, or of the FFLO phase in spin
polarized superconductors and CDWs.

\bigskip
\textbf{II.1. New routs to topological excitations}
\medskip

New interest to solitons in electronic processes emerges from a
discovery of the ferroelectric charge ordering in organic conductors
(Monceau et al., Brown et al. (2001), see \cite{lebed-springer} -
reviews \cite{brazov-springer,monceau}) and from nano-scale
experiments on internal tunneling in CDW materials (see
\cite{latyshev-prl:05,latyshev-prl:06} and short reviews
\cite{tunnel-ecrys:05}). The charge ordering allows us to observe
several types of solitons in conductivity, and solitons' bound pairs
in optics. The observed internal tunneling of electrons in CDWs goes
through the channel of amplitude solitons (\cite{latyshev-prl:05},
see below), which correspond to the long sought quasi-particle, the
spinon. Moreover, the resolved tunneling in the normally forbidden
subgap region \cite{latyshev-prl:06,tunnel-ecrys:05} recovers
collective quantum processes like coherent phase slips, $2\pi$
instantons. The same experiment gives an access to the reversible
reconstruction of the junction via spontaneous creation of a special
lattice of embedded $2\pi$ solitons, a grid of dislocations
\cite{latyshev-prl:06,tunnel-ecrys:05}.

On this solid basis we can extend the theory of solitons in quasi 1D
systems to arrive at a picture of combined topological excitations
in general strongly correlated systems:  from nearly
antiferromagnetic oxides to high gap superconductors. To extend
physics of solitons to the higher-D world, the most important
problem is the effect of confinement: as topological objects
connecting different degenerate vacuums, the solitons at D>1 acquire
an infinite energy unless they reduce or compensate their
topological charges. The problem is generic to all solitons, but it
becomes particularly interesting at the single electronic level,
where the spin-charge reconfinement appears as the result of
topological constraints. Especially interesting is the important
case of coexisting discrete and continuous symmetries. As a result
of their interference, the topological charge of solitons originated
by the discrete symmetry can be compensated by gapless degrees of
freedom originated by the continuous one. This is the scenario we shall
discuss through the rest of the article. Details and references can
be found in \cite{braz:00,braz-moriond,kirova}.

\bigskip
\textbf{II.2. Amplitude solitons with phase wings in quasi-1D spin-gap systems:
 CDWs and superconductors.}
\medskip

 Difference of states with even and odd numbers
of particles is a common issue for correlated electrons and
mesoscopics. Thanks to solitons, in an incommensurate CDW (ICDWs, an
arbitrary CDW wave number $Q$) it also shows up in a spectacular way
(S.B. 1978-80, see \cite{SB:84,SB:89}).
 \newline 1. Additional pair of electrons or holes is accommodated to the extended ground
state, for which the overall phase difference becomes $\pm2\pi$.
Phase increments are produced by phase slips, which provide the
spectral flow from the upper $+\Delta_{0}$ to the lower
$-\Delta_{0}$ rims of the single particle gap $2\Delta_{0}$. The
phase slip requires for the CDW amplitude $A(x,t)$ to pass through
zero, at which moment the complex order parameter has a shape of the
amplitude soliton (AS) - the kink,
 $A(x=-\infty)\leftrightarrow -A(x=+\infty$).
 \newline 2. This instantaneous configuration of the AS (Fig.1) becomes the stationary state
 for the case when only one electron is added to the system, or when the total spin
polarization is controlled to be nonzero. The AS carries the singly
occupied mid-gap state, thus having a spin $1/2$, but its charge is
compensated to zero by local dilatation of singlet vacuum states
\cite{SB:84,SB:89} - the AS is a realization of the "spinon".

As a nontrivial topological object ($O_{cdw}=A(x)\cos[Qx+\varphi]$
does not map onto itself), the pure AS is prohibited in $D>1$
environment. Nevertheless, the AS becomes allowed if it acquires
phase tails with the total increment $\delta \varphi=\pi$. The
length of these tails $\xi_{\varphi}\gg\xi_{0}=\hbar v_F/Delta_0$ is
determined by the weak interchain coupling. As in 1D, the sign of
$A(x)$ changes within the scale $\xi_{0}$ but further on, at the
scale $\xi_{\varphi}$, the factor $\cos[Qx+\varphi]$ also changes
the sign, thus leaving the product in $O_{cdw}$ to be invariant. As
a result, the 3D allowed particle is formed with the AS core
$\xi_{0}$ carrying the spin, and the two $\pi/2$-twisting wings
stretched over $\xi_{\varphi}$, each carrying the charge $e/2$. This
picture can be directly reformulated for quasi-1D superconductors by
redefining the meaning of the phase. The phase tails form now the
elementary $\pi$-junction \cite{braz-moriond,kwon:02}.

\bigskip
\textbf{II.3. A hole in the AFM environment.}
\medskip

Consider the quasi-1D system with repulsion at a nearly half filled
band, which is the SDW rout to a general doped antiferromagnetic
(AFM) Mott-Hubbard insulator (see \cite{braz:00} for more details).
The 1D bosonized Hamiltonian can be written schematically as
\[
H \sim\{C_{c}(\partial\varphi)^{2}-U\cos(2\varphi)\}+(C_{s}\partial \theta)^{2}
\]
Here $\varphi$ is the analog of the CDW displacement phase, $\theta$
is the spin rotation angle and $U$ is the Umklapp scattering
amplitude. In 1D, the excitations are the \textit{(anti)holon} as
the $\pm\pi$ soliton in $\varphi$, and the spin sound in $\theta$,
which are decoupled. The $\pi-$ solitons have been clearly
identified experimentally both in conductivity and optics (see
\cite{brazov-springer} for a review and interpretations). At D>1,
below the SDW ordering transition, the order parameter (the
staggered magnetization $<S_x+iS_y>$) is
 $O_{sdw}\sim\cos\varphi\exp(i\theta)$.
 To survive in $D>1$, the $\pi-$ soliton in $\varphi$
 ($\cos\varphi\rightarrow$ - $\cos\varphi$)
 should enforce the $\pi$ rotation in $\theta$,
 then the sign changes in both of the two factors composing $O_{sdw}$, cancel
each other and the configuration becomes allowed.

This construction can be generalized beyond quasi-1D systems by
considering a vortex configuration bound to an unpaired electron.
Extending to AFMs like the $\mathrm{CuO_2}$ planes, the SDW becomes
a staggered magnetization; the soliton becomes a hole, which motion
leaves the string of reversed AFM sublattices; the $\pi$- wings
become the magnetic semi-vortices. The resulting configuration is a
half-integer vortex ring of the staggered magnetization (a semi
roton) with the holon confined in its center (see  Fig.2  - in 2D,
the vortex ring is reduced to the pair of vortices.) Such combined
semi-vortices might be significant in sliding incommensurate SDWs
\cite{kirova}.

\bigskip
\textbf{II.4. Nucleus stripes as elementary excitations.}
\medskip

 The picture of combined topological excitations is well
established theoretically, as well as confirmed experimentally, for
weakly interacting chains. Since the results are symmetrically and
topologically characterized, they can be extrapolated qualitatively
to isotropic systems with strong coupling, where a clear microscopic
derivation is not available. Here the hypothesis is that, instead of
normal carriers excited or injected above the gap, the lowest states
are the symmetry broken configurations described above as a
semiroton-holon complex (spinon instead of holon for spin-gap
cases).

Strong interactions are necessary for this extrapolation from
quasi-1D. Stability of complex excitations is ultimately related to
local robustness of a stripe phase, which must sustain a
fragmentation due to quantum or thermal melting. Then any
termination point of one stripe (the dislocation within the regular
pattern) will be accompanied by the semiroton in accordance with the
quasi-1D picture: the combined soliton becomes a minimal element
(the nucleolus) of the melted stripe.

This picture allows us to answer the long staying question: are there
stripes present when we do cannot visualize them, e.g. in cuprates
away from the magic concentration $x=1/8$. The answer is that they
are still around - dispersed as an ensemble of their nucleuses.
These particles determine most of observable properties, usually
ascribed to conventional electrons. Thus, the associated mid-gap
states will give rise to observed growing arcs of Fermi
surfaces.Individual solitons may be better accessed when they are
trapped by dopants; that might be responsible for observation of
rod-like local structures in recent STM experiments in cuprates
{\cite{davis:06}, and more generally, for the whole issue of the
strong inhomogeneity \cite{bianconi:06}.

\bigskip
Details and illustrations can be found at the web site
\hfil \newline
\texttt{www.lptms.u-psud.fr/membres/brazov/seminars.html}

{Support is acknowledged by the INTAS grant 7972.}

\begin{figure}[htb]
\includegraphics*[width=7cm,height=4cm]{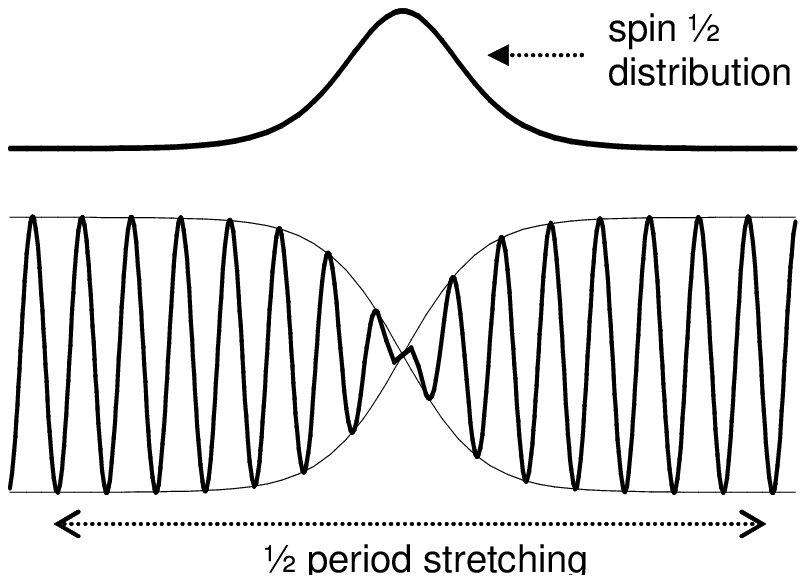}
\caption{Profiles for the amplitude soliton in ICDW.}%
\label{fig1}%
\end{figure}

\begin{figure}[htb]
\includegraphics[width=6cm,height=4cm]{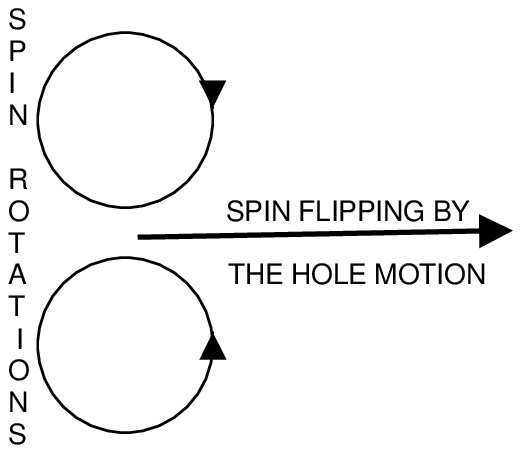} \caption{ Motion
of the kink-roton complex. For SDW or AFM, the string of the
amplitude reversal of the order parameter created by the holon is
cured by the semi-vortex pair (the loop in 3D) of the staggered
magnetization circulation. For ICDW or the superconductor, the
amplitude kink is provided by the spinon. For the ICDW the curls are
displacements contours for the half integer dislocation pair. For
the superconductor, the curls are lines of electric currents
circulating through the normal core carrying the unpaired spin.}
\label{fig2}
\end{figure}

\end{document}